\documentclass[sigconf]{acmart}
\usepackage{amsmath,amssymb,amsfonts}
\usepackage{graphicx}
\usepackage{xcolor}
\usepackage{booktabs} 
\usepackage{array, makecell} 
\usepackage[font={small},skip=5pt]{caption} %
\newcommand{\quotes}[1]{``#1''} 
\usepackage{msc} 
\usepackage{multirow} 

\usepackage{url}

\usepackage{breakurl}
\usepackage{hyperref}
\hypersetup{breaklinks=true}
\usepackage[compact]{titlesec} 
\usepackage{mathptmx} 
\usepackage{adjustbox} 
\usepackage{paralist} 
\usepackage{sistyle} 
\SIthousandsep{,} 

\newcommand*\justify{%
	\fontdimen2\font=0.4em 
	\fontdimen3\font=0.2em 
	\fontdimen4\font=0.1em 
	\fontdimen7\font=0.1em 
	\hyphenchar\font=`\- 
}

\newcommand{\var}[1]{\mathit{#1}} 

\begin{document}
	
\title{Does \quotes{www.} Mean Better Transport Layer Security?}
 
\author{Eman Salem Alashwali}
\affiliation{
	\institution{University of Oxford}
	\city{Oxford}
	\country{United Kingdom}\\
	\institution{King Abdulaziz University (KAU)}
	\city{Jeddah}
	\country{Saudi Arabia}}
\email{eman.alashwali@cs.ox.ac.uk}

\author{Pawel Szalachowski}
\affiliation{
	\institution{Singapore University of Technology and Design (SUTD)}
	\city{Singapore}
	\country{Singapore}}
\email{pawel@sutd.edu.sg}

\author{Andrew Martin}
\affiliation{
	\institution{University of Oxford}
	\city{Oxford}
	\country{United Kingdom}}
\email{andrew.martin@cs.ox.ac.uk}

\renewcommand{\shortauthors}{Alashwali, et al.}
 
\begin{abstract}
Experience shows that most researchers and developers tend to treat plain-domains (those that are \textit{not} prefixed with \quotes{www} sub-domains, e.g. \quotes{example.com}) as synonyms for their equivalent www-domains (those that are prefixed with \quotes{www} sub-domains, e.g. \quotes{\justify www.example.com}). In this paper, we analyse datasets of nearly two million plain-domains against their equivalent www-domains to answer the following question: \textit{Do plain-domains and their equivalent www-domains differ in TLS security configurations and certificates? If so, to what extent?} Our results provide evidence of an interesting phenomenon: plain-domains and their equivalent www-domains differ in TLS security configurations and certificates in a non-trivial number of cases. Furthermore, www-domains tend to have stronger security configurations than their equivalent plain-domains. Interestingly, this phenomenon is more prevalent in the most-visited domains than in randomly-chosen domains. Further analysis of the top domains dataset shows that 53.35\% of the plain-domains that show one or more weakness indicators (e.g. expired certificate) that are not shown in their equivalent www-domains perform HTTPS redirection from HTTPS plain-domains to their equivalent HTTPS www-domains. Additionally, 24.71\% of these redirections contains plain-text HTTP intermediate URLs. In these cases, users see the final www-domains with strong TLS configurations and certificates, but in fact, the HTTPS request has passed through plain-domains that have less secure TLS configurations and certificates. Clearly, such a set-up introduces a weak link in the security of the overall interaction.
\end{abstract}

\copyrightyear{2019}
\acmYear{2019}
\setcopyright{acmcopyright}
\acmConference[ARES '19]{Proceedings of the 14th International Conference on Availability, Reliability and Security (ARES 2019)}{August 26--29, 2019}{Canterbury, United Kingdom}
\acmBooktitle{Proceedings of the 14th International Conference on Availability, Reliability and Security (ARES 2019) (ARES '19), August 26--29, 2019, Canterbury, United Kingdom}
\acmPrice{15.00}
\acmDOI{10.1145/3339252.3339277}
\acmISBN{978-1-4503-7164-3/19/08}


\maketitle

\section{Introduction}
\label{sec:intro}
The \quotes{www.} (world-wide web) prefix has become a de facto standard for domains running websites. Plain-domains (those without the \quotes{www.} prefix, e.g. \quotes{example.com}) are usually treated as synonyms for their equivalent www-domains (those that are prefixed with \quotes{www.}, e.g. \quotes{www.example.com}). \par 
As a concrete example, from an application development perspective, recently, in August 2018, Google's \texttt{Chrome}\footnote{As a shorthand, we use the term \texttt{Chrome} for the rest of the paper to refer to Google's \texttt{Chrome}.} version 69.0.3497.81 decided to hide the \quotes{www} and \quotes{m} (\quotes{m} for mobile) sub-domains from
the steady-state URL in \texttt{Chrome}'s address bar by default~\cite{nichols18}. Google describes these sub-domains as \quotes{trivial} in \texttt{Chrome}'s custom settings, where users can enable displaying these sub-domains through the
following URI: \quotes{\url{
chrome://flags/\#omnibox-ui-hide-steady-state-url-scheme-and-sub-domains}}. This
change by Google has received criticism and media attention from the
security community, as shown in~\cite{nichols18}\cite{liam18}\cite{shawn18}. One of the reasons for not welcoming this change is that it can
cause confusion~\cite{nichols18}. For example, two addresses, one with a \quotes{www} sub-domain and another without a \quotes{www} sub-domain can point to completely different websites~\cite{nichols18}. One user reported \texttt{Chrome}'s new behaviour of
hiding the \quotes{www} and \quotes{m} sub-domains as a bug~\cite{chromium18}. In the same report~\cite{chromium18}, another user has pointed out that the \texttt{Safari} mobile browser also hides the \quotes{www} sub-domains from its address bar~\cite{chromium18}. Subsequently, other users have reported that the Google search engine sometimes hides the \quotes{www} sub-domains in search results and uses plain-domains format~\cite{lawrence18}. However, we are unable to reproduce the issue (bug) reported in~\cite{chromium18} regarding \texttt{Chrome}'s new behaviour since \texttt{Chrome} does not provide an archive for old versions. Nevertheless, the reported bug includes several confirming responses~\cite{chromium18}, in addition to media reports, such as~\cite{nichols18}\cite{liam18}\cite{shawn18}, which provide sufficient evidence that the issue has existed in that particular version, either in the desktop, or the mobile version, or both. Our test of \texttt{Chrome}'s latest version shows that subsequent versions (as of March 2019, version 73.0.3683.86) have reverted this change from a default to a custom setting. \par 

From a research perspective, in particular, on Internet measurement studies which are concerned with examining domains' TLS security configurations and certificates, we observe different treatments for the examined domain names. A considerable number of these studies make use of the \texttt{Alexa} top domains list~\cite{alexa18}. \texttt{Alexa} represents domains as plain-domains, except for a small percentage\footnote{This percentage is around 4.99\% in our \texttt{Alexa} dataset.} of domains that include sub-domains including \quotes{www} sub-domains. Some studies, such as~\cite{li17}, add the \quotes{www} sub-domains to the examined domains, and report the results based on TLS handshakes with www-domains. Other studies, such as~\cite{huang14}, first try to establish a TLS handshake with the examined domain \quotes{as is}, and if the handshake with the domain \quotes{as is} has failed, they make a second handshake with the domain's equivalent www-domain, and report the results. The third line of Internet measurement studies such as~\cite{durumeric13}\cite{felt17}, do not mention adding any \quotes{www} sub-domains to the examined domains, hence, we assume that they treat the list's domains (e.g. the \texttt{Alexa} list in these studies) \quotes{as is}. However, it remains unclear whether plain-domains differ from their equivalent www-domains in terms of TLS configurations and certificates? If one type of domains, e.g. www-domains, provides better TLS security configurations than their equivalent plain-domains, or vice versa, to what extent does taking one approach over another affect the overall results? In particular, regarding the domains' adoption of TLS security configurations, e.g. protocol versions, or the domains' certificates status results.  \par  

In this paper, we try to answer the following question: \textit{Do plain-domains and their equivalent www-domains differ in TLS security configurations and certificates? If so, to what extent?}

\section{Background}
\label{sec:background}
The Transport Layer Security (TLS) protocol~\cite{rescorla18tls13}\cite{rescorla08tls12}, formerly known as Secure Socket Layer (SSL), is one of the most important and widely used security protocols to date. The latest version of TLS is known as TLS~1.3~\cite{rescorla18tls13}. We refer to versions prior to TLS~1.3 (in particular, TLS~1.2, TLS~1.1, and TLS~1.0) as pre-TLS~1.3. TLS operates below application layer protocols, to provide data confidentiality, integrity, and authentication. TLS consists of multiple sub-protocols, including the handshake protocol. In the handshake protocol, both communicating parties authenticate each other, and negotiate security-sensitive parameters, including the protocol version and the ciphersuite, that will be used to secure subsequent messages of the protocol. The ciphersuite is an identifier, represented by a string or a hexadecimal value, which defines the cryptographic algorithms (e.g. symmetric encryption algorithm) and their parameters (e.g. key length) that will be used in subsequent messages of the protocol. In pre-TLS~1.3 versions, the ciphersuite defines the key-exchange, authentication, symmetric encryption, and hash algorithms. However, in TLS~1.3, the key-exchange is separated from the ciphersuites, and it is now negotiated in specific extensions. \par 

The version and ciphersuite choices are intrinsic in determining the security guarantees that the protocol can provide in a particular session. Some ciphersuites provide stronger security guarantees than others. For example, Forward Secrecy (FS) guarantees that a compromise in a server's long-term private-key does not compromise past session keys~\cite{menezes96}. Similarly, Authenticated Encryption (AE) provides confidentiality, integrity, and authenticity, simultaneously, which provides stronger resilience against attacks over the MAC-then-encrypt schemes~\cite{vaudenay02}\cite{alfardan13}. The same applies to the protocol version. Each version of the protocol prevents attacks discovered in previous versions, and provides more security guarantees. For example, TLS~1.3 enforces FS key-exchange algorithms and AE ciphersuites (FS+AE) by design. TLS~1.2 supports AE in addition to FS but both are optional, while TLS~1.1 and TLS~1.0 do not support AE ciphersuites. \par 

\begin{figure}[!tp]
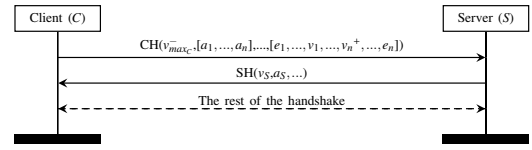

	\centering
	\vspace{-20pt}
	\resizebox{0.85\columnwidth}{!}{
		\setmsckeyword{} 
		\drawframe{no} 
		\begin{msc}[large values, /msc/level height=0.6cm, /msc/label 
		distance=0.5ex , /msc/first level height=0.6cm, /msc/last level 
		height=0.6cm, /msc/top head dist=0, /msc/bottom foot dist=0, 
		/msc/environment distance=0, /msc/foot height=1.5ex]{}
			\setlength{\instwidth}{2\mscunit} 
			\setlength{\instdist}{8\mscunit} 
			
			\declinst{C}{}{Client ($C$)}
			\declinst{S}{}{Server ($S$)}
			
			\mess{CH($v_{\mathit{max}_C}^-$,[$a_1,...,a_n$],...,[$e_1,...,{[v_1,...,v_n]}^+,...,e_n$])} {C}{S}
			\nextlevel
			
			\mess{SH($v_S$,$a_S,...$)} {S}{C}
			\nextlevel
			
			\mess*{The rest of the handshake}{C}{S}
			\mess*{}{S}{C}
			
		\end{msc}
	} 
	\vspace{-10pt}
	\caption{Illustration of the TLS version and ciphersuite negotiation. Parameters followed by \quotes{$-$} are deprecated in TLS~1.3 but still included for backward compatibility, while those followed by \quotes{$+$} are newly introduced in TLS 1.3. The unmarked parameters are mutual to both versions.}
	\label{fig:TLS-CH} 	
\end{figure}

As depicted in \autoref{fig:TLS-CH}, at the beginning of a new TLS handshake, the client sends a \texttt{ClientHello} (\texttt{CH}) message to the server. The \texttt{CH} contains several security-sensitive parameters, including: first, the client's supported versions. In pre-TLS~1.3, this parameter is sent as a single value $v_{\mathit{max}_C}$. In TLS~1.3, it is sent as a list [$v_1,...,v_n$] in the \quotes{supported\_versions} extension. The $v_{\mathit{max}_C}$ is still included in TLS~1.3 \texttt{CH} for backward compatibility, and its value is set to TLS~1.2. Second, the client's supported ciphersuites is sent as a list [$a_1,...,a_n$]. Third, the client's supported extensions as a list [$e_1,...,e_n$] at the end of the message. In TLS~1.3, the extensions must at least include the \quotes{supported\_versions}, while in TLS~1.2, the extensions are optional. Upon receiving the \texttt{CH}, the server selects a single mutually supported version $v_S$ and ciphersuite $a_S$ from the client's offer. The server then responds with a \texttt{ServerHello} (\texttt{SH}) containing $v_S$ and $a_S$. The server's selection of $v_S$ and $a_S$ is imposed on the client. That is, the server's decision on the version and ciphersuite is the final one. If the server does not support the client's offered versions or ciphersuites, the server responds with a handshake failure alert. 

\section{Dataset}
\label{sec:dataset}
Our study includes two datasets: \texttt{top-domains}, which contains \num{829873} distinct most-visited domains globally, and \texttt{\justify random-domains}, which contains \num{992422} distinct random domains. We now define some of the terms that we use throughout the paper: \begin{inparaenum} 
\item \textbf{Main-domains:} domains consisting of a Top Level Domain (TLD) (e.g. \quotes{com}) prefixed by a single label, and do not have any further sub-domains, e.g. \quotes{example.com}. 
\item \textbf{Plain-domains:} domains that are not prefixed with \quotes{www} sub-domains, e.g. \quotes{example.com}. 
\item \textbf{www-domains:} domains that are prefixed with \quotes{www} sub-domains, e.g. \quotes{www.example.com}.
\end{inparaenum} Our scope is limited to general TLDs (gTLDs), i.e. single-level TLDs, such as \quotes{com}, and does not include multi-level TLDs, such as country-code TLDs (ccTLDs), for example \quotes{ac.uk}. This is to avoid the complexity of distinguishing domains that have sub-domains from domains that have ccTLDs which is somewhat difficult to achieve with 100\% accuracy. In what follows we describe how we build and pre-process each dataset. 

\subsection{Top-Domains}
The \texttt{top-domains} dataset is derived 
from the \texttt{Alexa} list of the top one million most-visited domains globally~\cite{alexa18}, retrieved on December 11, 2018. Since our study targets plain-domains initially, from the 1 million domains, we extract the domains that are classified as plain-domains and also classified as main-domain (see the previous paragraph for our definitions of plain-domains and main-domains). We end up with \num{829873} distinct top domains. To create their equivalent www-domains, we make a replica of the extracted plain-domains, then we prefix each domain with \quotes{www.}. 

\subsection{Random-Domains}
Our \texttt{random-domains} dataset initial size is one million domains, 
extracted from a large dataset of \num{54063220} domains that have 
successfully completed TLS handshakes in Amann et al.~\cite{amann17}. The domains in~\cite{amann17} are collected from multiple sources including other lists and previous work. We pre-process our \texttt{random-domains} dataset in two steps: First, to maintain consistency with the \texttt{top-domains} dataset format, from Amann et al.'s list~\cite{amann17}, we extract one million random domains that are classified as main-domains and plain-domains. Second, from the just extracted domains, we exclude the domains that already exist in the \texttt{top-domains} dataset, to avoid repetition. We finish with \num{992422} distinct random domains. Finally, to create their equivalent www-domains, we create a replica of the dataset, then we prefix each domain with \quotes{www.}.

\section{Methodology}
\label{sec:methodology}
\subsection{Data Collection}
To collect data, we run a TLS client that takes the domain names from our datasets as input, performs a TLS handshake with each domain, and outputs the handshake's response data. For each dataset, namely the \texttt{top-domains} and \texttt{random-domains} datasets, the client performs TLS handshakes with plain-domains and their equivalent www-domains concurrently, using two separate TLS client instances. We perform two types of TLS handshakes per domain: one utilising TLS~1.2 client configurations, and the second utilising TLS~1.3 client configurations. Our clients' configurations (mainly, the supported TLS versions and ciphersuites) are based on \texttt{Chrome}'s latest version configurations. We choose to base our clients' TLS configurations on \texttt{Chrome} because it is the most representative TLS client on the Internet today. As of February 2019, \texttt{Chrome}'s usage is 79.7\%~\cite{w3schools18}. To increase our confidence in the obtained results, we run the above-described experiment twice. The first experiment run from the 11\textsuperscript{th} to 12\textsuperscript{th} December 2018, and the second one run from the 5\textsuperscript{th} to 6\textsuperscript{th} March 2019.\par 

The TLS handshakes' responses, which include the servers' TLS versions, ciphersuites, and certificates are first stored in \texttt{json} format, parsed, then stored and analysed using the \texttt{MySQL} database and queries. In terms of counting the servers' responses, we follow a best-effort approach. That is, we count the responding domains but we do not investigate the reasons for the non-responding ones. However, we eliminate the possibility of servers' rate-limit\footnote{Rate-limit is a technique used to mitigate Denial of Service (DoS) attacks by specifying a threshold for the incoming requests by a particular source address.} as a reason for the non-responding domains, as our client performs only two handshakes (TLS~1.2 and TLS~1.3) per domain type (plain-domains and www-domains) per dataset (\texttt{top-domains} and \texttt{random-domains}). We also eliminate the time gap\footnote{By time gap we refer to a situation where a domain can be alive in one handshake (e.g. with plain-domain), but goes down in the second handshake (e.g. with www-domain).} as a reason for the non-responding domains, as our clients' handshakes with both plain-domains and their equivalent www-domains are performed concurrently. In our analysis, we only consider the domains that responded to both plain-domains and their equivalent www-domains handshakes. \par 

In terms of the TLS client that performs the TLS handshakes, our TLS client is based on \texttt{tls-scan}~\cite{tls-scan16}, an open source fast TLS scanner capable of performing concurrent TLS connections. We then customise the \texttt{tls-scan} to utilise the \texttt{OpenSSL 1.1.0g} TLS library for our TLS~1.2 client and the \texttt{OpenSSL 1.1.1a} for our TLS~1.3 client. We customise our clients' offered TLS versions and ciphersuites to be equivalent to those offered by \texttt{Chrome}'s latest version for our TLS~1.3 client, and \texttt{Chrome}'s latest version pre-TLS~1.3 configurations for our TLS~1.2 client. The Server Name Indication (SNI)\footnote{The SNI is a TLS extension that passes the domain name in the TLS handshake in order to get more accurate responses in virtual hosting environments.} extension is included by default. We set the concurrency argument to 50 and 25 concurrent connections (these vary between the first and second experiment), and the timeout argument is set to 5 seconds. We run the experiments at the University of Oxford, on computers equipped with 1000 Mbps wired Ethernet cards. \par 

After the TLS scan is complete, to better understand the results, we conduct an HTTPS redirection scan. We build a multi-threaded redirection scanner based on the Python's \texttt{Requests} library~\cite{requests19}, which provides HTTPS redirection data. The scanner takes as input the domains that have provided responses from both plain-domains and their equivalent www-domains in the TLS~1.3 handshakes in the second TLS scan experiment. The scanner then performs an HTTPS request with each domain without certificate validation, and collects the HTTPS redirection chain from the initial until the final URL (or domain) for each request. We then store and analyse the data using the \texttt{MySQL} database and queries. The redirection scan run from the 15\textsuperscript{th} to 17\textsuperscript{th} March 2019 at the University of Oxford. The short time span between the last TLS scan experiment and the redirection scan (around 8 days) allows us to correlate the redirection behaviour to the TLS configurations and certificate behaviour with confidence.    

\subsection{Ethical Considerations}
First, we do not collect any private data. The data we collect are public metadata. Second, we do not perform an exhaustive number of handshakes on any single server. Our client's handshakes can by no means be classified as a Denial of Service (DoS) attack. Third, we use a designated public IPv4 address for the scanning device instead of Network Address 
Translation (NAT), to avoid potential disturbance to other users in our organisation's network, if a server has blocked our scanning device's IP. Fourth, we use an informative DNS name for our scanning device that contains \quotes{TLS probing}, to help server administrators identify our device's activity in their logs. Finally, we inform our University's IT and Security teams, so they expect a high volume of outgoing connections from our experiment devices, and to expect some incoming blacklisted certificates from random servers. 

\subsection{Data Analysis}\label{sec:analysis}
To answer the first part of our research question (\textit{do
plain-domains and their equivalent www-domains differ in TLS security
configurations and certificates?}), we count the number of domains that: \begin{inparaenum}[(a)] \item responded to  both
plain-domains and their equivalent www-domains TLS handshakes, and \item provide different values for one or more of the following TLS configurations: \end{inparaenum} 
\begin{inparaenum}
	\item TLS versions,
	\item Ciphersuites,
	\item Versions \textit{and} ciphersuites,
	\item Versions \textit{or} ciphersuites,
	\item Leaf certificate Subject Key Identifiers (SKI), which is an X509 extension that provides the means for identifying certificates with a particular public-key~\cite{cooper08}, 
	\item Leaf certificate fingerprints (SHA-1), which provide the means for identifying certificates in general (including all the certificate's fields),
	\item IPv4, although not directly a TLS security configuration, but we include it as a useful general result.
\end{inparaenum}  

To answer the second part of our research question (\textit{If so, to what extent?}), we first define some weakness indicators denoted by $\var{weak}$. Then, we analyse plain-domains and their equivalent www-domains against $\var{weak}$. In what follows, we list these weakness indicators, and define when they are satisfied by domains: 

\begin{compactenum}
\item v.$<$TLS~1.3: is satisfied by domains that select TLS versions less than TLS~1.3 (the latest version). 

\item v.$<$TLS~1.2: is satisfied by domains that select TLS versions less than TLS~1.2. Versions less than TLS~1.2 are officially weak and should not be used today.

\item non-FS: is satisfied by domains that select non-FS ciphersuites (despite the AE or any other properties). We define non-FS ciphersuites\footnote{All our definitions are based on \texttt{Chrome}'s configurations.} as those that do not support the Elliptic-Curve Diffie-Hellman (ECDHE), and are also not negotiated with version TLS~1.3 since TLS~1.3 enforces FS by design in a separate extension. Non-FS ciphersuites provide fewer security guarantees than FS ciphersuites.
    
\item non-FS+non-AE: is satisfied by domains that select non-FS+non-AE ciphersuites, i.e. neither provide FS nor AE. We define non-FS+non-AE ciphersuites as those that do not support ECDHE, are not negotiated with version TLS~1.3, do not support the ChaCha20 symmetric encryption, and do not support the GCM symmetric encryption mode. This indicator is not satisfied by domains that select FS+AE, i.e. those that either support ECDHE or negotiated with version TLS~1.3, and support either the ChaCha20 symmetric cipher or the GCM mode. Non-FS+non-AE ciphersuites provide fewer security guarantees than those that provide both FS and AE (FS+AE).

\item Exp. Certs.: is satisfied by domains that provide expired certificates. The expiration check is performed by the \texttt{tls-scan} client against the scan date~\cite{tls-scan16}.

\item Invalid Certs.: is satisfied by domains that provide invalid certificates. The certificate validation is performed by the \texttt{\justify tls-scan} client using the \texttt{\justify SSL\_get\_verify\_result} function in the \texttt{\justify OpenSSL} library against our updated \texttt{\justify Ubuntu~18.04} certificates store. It validates the certificate's signature, trust chain, and other verification steps specified in~\cite{verify18}. 

\item Key$<$2048: is satisfied by domains that use RSA certificates with a key length of less than 2048-bit. The minimum recommended RSA key length today is 2048-bit\footnote{https://www.keylength.com.}.
\end{compactenum}

Then, we count the domains' responses under the following conditions, where $\var{weak}$ denotes a weakness indicator, $\var{plain}$ denotes plain-domains that have responses for their equivalent www-domains, and $\var{www}$ denotes www-domains that have responses for their equivalent plain-domains:
\begin{compactenum}
\item $\var{weak}_{\var{plain}}$: $\var{weak}$ is satisfied by $\var{plain}$.
\item $\var{weak}_{\var{plain}} \wedge\lnot \var{weak}_{\var{www}}$:  $\var{weak}$ is satisfied by $\var{plain}$ and $\var{weak}$ is \textit{not} satisfied by $\var{www}$.
\item $\var{weak}_{\var{www}}$: $\var{weak}$ is satisfied by $\var{www}$.
\item $\var{weak}_{\var{www}} \wedge\lnot \var{weak}_{\var{plain}}$:  $\var{weak}$ is satisfied by $\var{www}$ and $\var{weak}$ is \textit{not} satisfied by $\var{plain}$.
\end{compactenum}

\section{Results}
\label{sec:results}
As described earlier in the methodology, we have conducted two experiments over a 3-month time span, to increase our confidence in the obtained results. In this section, we only report the results from the latter experiment (March 2019). However, they are consistent with the former experiment (December 2018).

\subsection{Responding Servers}
\autoref{tab:responses_scan2} summarises the number of successful TLS handshake responses from plain-domains and their equivalent www-domains in the \texttt{top-domains} and \texttt{random-domains} datasets, in both types of client handshakes, TLS~1.2 and TLS~1.3. Responses from top domains are higher than those from random domains. This is mostly due to the fact that the TLS adoption rate in top domains is higher than that in random domains. Furthermore, the \texttt{top-domains} dataset is built from a recent \texttt{Alexa} list which contains active domains, while the \texttt{random-domains} dataset is built from a less recent list from Amann et al.~\cite{amann17}, which may contain many inactive domains. We notice that in the \texttt{top-domains} dataset, www-domains responses are higher than plain-domains responses by 2.3\%. This means that top domains tend to have more active www-domains than plain-domains. On the other hand, in the \texttt{random-domains} dataset, both plain-domains and www-domains responses are nearly equal, with slightly more responses from plain-domains than www-domains. 

\begin{table}[!tp]
	\centering
	\caption{Responding servers to our TLS clients' handshakes.}
	\label{tab:responses_scan2} 
	\begin{adjustbox}{max width=0.9\columnwidth}
		\begin{tabular}{llllr@{\hspace{5pt}}r}
			\toprule
			\thead{Client}& \thead{Dataset} & \thead{Size} &\thead{Type} & \multicolumn{2}{r}{\thead{Responses}} \\
			\midrule
			\multirow{4}{*}{\texttt{TLS~1.2}}&
			\multirow{2}{*}{\texttt{top-domains}} & \multirow{2}{*}{\num{829873}} 
			& plain & \num{691200}&(83.29\%)\\
			\cline{4-6}
			& & & www  & \num{710509} & (85.62\%)\\ 
			\cline{2-6}
			
			& \multirow{2}{*}{\texttt{random-domains}} & \multirow{2}{*}{\num{992422}}	   
			& plain & \num{623869}&(62.86\%)\\
			\cline{4-6}
			& & & www  & \num{620884}& (62.56\%) \\
			
			\midrule
			\multirow{4}{*}{\texttt{TLS~1.3}}&
			\multirow{2}{*}{\texttt{top-domains}} & \multirow{2}{*}{\num{829873}} 
			& plain & \num{691145}& (83.28\%)\\
			\cline{4-6}
			& & & www  & \num{710496}& (85.62\%)\\ 
			\cline{2-6}
			
			& \multirow{2}{*}{\texttt{random-domains}} & \multirow{2}{*}{\num{992422}}	   
			& plain & \num{622904} & (62.77\%)\\
			\cline{4-6}
			& & & www  & \num{620062} & (62.48\%) \\
			\bottomrule
		\end{tabular}
	\end{adjustbox}
\end{table}

\begin{table*}[!tp]
	\centering
	\caption{Differences between plain-domains and their equivalent www-domain against some TLS configurations. The \quotes{Responses} column represents the number of domains that responded to \textit{both} plain-domain and their equivalent www-domain TLS handshakes.}
	\label{tab:diffs_scan2}	
	\begin{adjustbox}{max width=\textwidth}
		\begin{tabular}{llrr@{\hspace{3pt}}rr@{\hspace{3pt}}rr@{\hspace{3pt}}rr@{\hspace{3pt}}rr@{\hspace{3pt}}rr@{\hspace{3pt}}rr@{\hspace{3pt}}r}
			\toprule
			& & & \multicolumn{14}{c}{Different}\\
			\cline{4-17}
			\thead{Client}&\thead{Dataset} &\thead{Responses} & \multicolumn{2}{c}{\thead{Version}} 
			& \multicolumn{2}{c}{\thead{Ciphersuite}} &  
			\multicolumn{2}{c}{\begin{tabular}{@{}c@{}}\thead{Version $\wedge$\\Ciphersuite} 
			\end{tabular}}& 
			\multicolumn{2}{c}{\begin{tabular}{@{}c@{}}\thead{Version 
						$\vee$\\Ciphersuite} \end{tabular}}
			& \multicolumn{2}{c}{\thead{SKI}} & \multicolumn{2}{c}{\thead{Cert.}} & \multicolumn{2}{c}{\thead{IPv4}}\\
			
			\midrule
			\multirow{2}{*}{\texttt{TLS~1.2}}& 
			\texttt{top-domains} & \num{678337}  
			& \num{1472}&(0.22\%) & \num{23600}&(3.48\%) & \num{1437}&(0.21\%)& \num{23635}&(3.48\%) & \num{74433}&(10.97\%)& \num{75359}&(11.11\%) & \num{125767}&(18.54\%) \\
			\cline{2-17}
			
			&\texttt{random-domains} & \num{614184} 
			& \num{748}&(0.12\%) & \num{7883}&(1.28\%) & \num{737}&(0.12\%) & \num{7894}&(1.29\%)& \num{52465}&(8.54\%)& \num{52844}&(8.60\%) & \num{57037}&(9.29\%)\\
			\midrule 
			
			\multirow{2}{*}{\texttt{TLS~1.3}}& 
			\texttt{top-domains} & \num{678214}  
			& \num{9187}&(1.35\%) & \num{24819}&(3.66\%) & \num{9151}&(1.35\%)& \num{24855}&(3.66\%)& \num{74375}&(10.97\%)& \num{75311}&(11.10\%) & \num{125861}&(18.56\%) \\
			\cline{2-17}
			
			&\texttt{random-domains} & \num{612839}  
			& \num{3292}&(0.54\%) & \num{8386}&(1.37\%) & \num{3282}&(0.54\%)& \num{8396}&(1.37\%) & \num{52318}&(8.54\%)& \num{52718}&(8.60\%) & \num{57063}&(9.31\%)\\
			\bottomrule
		\end{tabular}
	\end{adjustbox}
\end{table*}


\begin{table*}[!tp]
	\centering
	\caption[]{Breakdown of some weakness indicators $\var{weak}$ in plain-domains and their equivalent www-domains based on the TLS~1.3 client handshake responses. The indentation in the \quotes{Type/Condition} column indicates that the percentages of the indented row's results are computed over the previous row's results. The percentages of the non-indented row's results are computed over the \quotes{Responses} values. Recall: $\var{weak}$ denotes weakness indicator; $\var{weak}_{\var{plain}}$ denotes $\var{weak}$ is satisfied by $\var{plain}$; $\var{weak}_{\var{plain}} \wedge\lnot \var{weak}_{\var{www}}$ denotes  $\var{weak}$ is satisfied by $\var{plain}$ and $\var{weak}$ is not satisfied by $\var{www}$; $\var{weak}_{\var{www}}$ denotes $\var{weak}$ is satisfied by $\var{www}$; and $\var{weak}_{\var{www}} \wedge\lnot \var{weak}_{\var{plain}}$ denotes  $\var{weak}$ is satisfied by $\var{www}$ and $\var{weak}$ is not satisfied by $\var{plain}$.}
	
	\label{tab:detailed_diffs_scan2}
	\begin{adjustbox}{max width=\textwidth}
		\begin{tabular}{lrlr@{\hspace{3pt}}rr@{\hspace{3pt}}rr@{\hspace{3pt}}rr@{\hspace{3pt}}rr@{\hspace{3pt}}rr@{\hspace{3pt}}rr@{\hspace{3pt}}r}
			\toprule
			\multirow{2}{*}{\thead{Dataset}} & \multirow{2}{*}{\thead{Responses}} & \multirow{2}{*}{\thead{Type/Condition}} &
			\multicolumn{14}{c}{Weakness Indicator ($\var{weak}$)}\\
			\cline{4-17}
			& & & \multicolumn{2}{c}{\thead{v.$<$TLS~1.3}} &
			\multicolumn{2}{c}{\thead{v.$<$TLS~1.2}} & \multicolumn{2}{c}{\thead{non-FS}} & \multicolumn{2}{c}{\thead{non-FS+non-AE}}  & \multicolumn{2}{c}{\thead{Exp. Cert.}} & \multicolumn{2}{c}{\thead{Invalid Cert.}} & \multicolumn{2}{c}{\thead{Key$<$2048}}\\
			\midrule
			
			\multirow{4}{*}{\texttt{top-domains}} 		   
			& \multirow{4}{*}{\num{678214}} 
			& $\var{weak}_{\var{plain}}$&
			\num{542267}&(79.96\%) & \num{15731}&(2.32\%) &  \num{27777}&(4.10\%) & 
			\num{17368}&(2.56\%) & \num{24521}&(3.62\%) & \num{55932}&(8.25\%) & \num{6306}&(0.93\%) \\
			\cline{4-17}
			& & \quad $\var{weak}_{\var{plain}} \wedge\lnot \var{weak}_{\var{www}}$& 
			\num{5280}&(0.97\%) & \num{1179}&(7.49\%) &  \num{3151}&(11.34\%) & 
			\num{1907}&(10.98\%) & \num{2510}&(10.24\%) & \num{5281}&(9.44\%) & \num{278}&(4.41\%) \\
			\cline{4-17}
			
			& & $\var{weak}_{\var{www}}$ & 
			\num{539543}&(79.55\%) & \num{14846}&(2.19\%) & \num{25184}&(3.71\%) & 
			\num{15710}&(2.32\%) & \num{23214}&(3.42\%) & \num{52773}&(7.78\%) & \num{6116}&(0.90\%) \\
			\cline{4-17}
			
			& & \quad $\var{weak}_{\var{www}} \wedge\lnot \var{weak}_{\var{plain}}$ &
			\num{2556}&(0.47\%) & \num{294}&(1.98\%) & \num{558}&(2.22\%) & \num{302}&(1.92\%) & \num{1203}&(5.18\%) & \num{2122}&(4.02\%) & \num{135}&(2.21\%)\\
			\midrule
			
			\multirow{4}{*}{\texttt{random-domains}}	   
			& \multirow{4}{*}{\num{612839}}  
			& $\var{weak}_{\var{plain}}$ &
			\num{545924}&(89.08\%) & \num{27774}&(4.53\%) &  \num{38653}&(6.31\%) & 
			\num{30621}&(5.00\%) & \num{55734}&(9.09\%) & \num{107521}&(17.54\%) & \num{18781}&(3.06\%) \\
			\cline{4-17}
			
			& & \quad $\var{weak}_{\var{plain}} \wedge\lnot \var{weak}_{\var{www}}$&
			\num{1667}&(0.31\%) & \num{536}&(1.93\%) &  \num{770}&(1.99\%) & 
			\num{496}&(1.62\%) & \num{911}&(1.63\%) & \num{1684}&(1.57\%) & \num{191}&(1.02\%) \\
			\cline{4-17}
			
			& & $\var{weak}_{\var{www}}$ &
			\num{545171}&(88.96\%) & \num{27448}&(4.48\%) & \num{38152}&(6.23\%) & 
			\num{30280}&(4.94\%) & \num{55350}&(9.03\%) & \num{106663}&(17.40\%) & \num{18628}&(3.04\%) \\
			\cline{4-17}
			
			& & \quad $\var{weak}_{\var{www}} \wedge\lnot \var{weak}_{\var{plain}}$&
			\num{914}&(0.17)\% & \num{210}&(0.77\%) & \num{269}&(0.71\%) & \num{209}&(0.69\%) & \num{527}&(0.95\%) & \num{826}&(0.77\%) & \num{46}&(0.25\%)\\
			\bottomrule
		\end{tabular}
	\end{adjustbox}
\end{table*}


\subsection{The Difference is in the Detail}\label{sec:diffs}
As depicted in \autoref{tab:diffs_scan2}, our results show that the client type, i.e. TLS~1.2 vs. TLS~1.3, does not make a noticeable effect in terms of the exhibited differences in TLS configurations and certificates between plain-domains and their equivalent www-domains. However, in the \quotes{version} difference, TLS~1.3 client shows a higher percentages (1.35\% and 0.54\%) than TLS~1.2 client (0.22\% and 0.12\%) in the \texttt{top-domains} and \texttt{random-domains} datasets, respectively. This can be explained by the fact that TLS~1.3 was standardised in 2018, which suggests that domain administrators may have updated the TLS version of one domain, e.g. the plain-domain, but not its equivalent, e.g. the www-domain, or vice versa. In terms of other TLS configurations and certificate differences at both clients TLS~1.3 and TLS~1.2, in general, top domains show higher percentages of differences in the TLS configurations and certificates between plain-domains and their equivalent www-domains than random domains. In general, the most significant differences are in the leaf certificate fingerprints (\quotes{Cert.}), certificate \quotes{SKI}, and in the selected \quotes{version or ciphersuite}. For example, as illustrated in \autoref{tab:diffs_scan2}, in the \texttt{top-domains} dataset, more than 11\% of plain-domains provide a different certificate fingerprint than their equivalent www-domains. Over 10\% of plain-domains provide a different SKI than their equivalent www-domains. Over 3.4\% select a different version or ciphersuite than their equivalent www-domains. On the other hand, the \texttt{random-domains} dataset shows lower percentages of differences (see \autoref{tab:diffs_scan2}).

\subsection{When \quotes{\texttt{www.}} Means Better TLS Security}\label{sec:www_security}
For further analysis, we compare plain-domains and their equivalent www-domains against several defined weakness indicators denoted by $\var{weak}$ (see \autoref{sec:analysis} for further details about the methodology). In this section, we limit our analysis to TLS~1.3 client handshake results as TLS~1.3 handshake is more representative of an updated TLS client today, such as web browsers (recall that in our TLS scan experiments, we perform two types of handshakes for each domain, one utilising TLS~1.3 configurations, and the second utilising TLS~1.2 configurations). As shown in \autoref{tab:detailed_diffs_scan2}, it is always the case that there are fewer www-domains that satisfy weakness indicators compared to plain-domains that do so (see \autoref{tab:detailed_diffs_scan2}, under the \quotes{Type/Condition} column, compare the results of row \quotes{$\var{weak}_{\var{plain}}$} to those of \quotes{$\var{weak}_{\var{www}}$} in each dataset). Also, it is always the case that the number of plain-domains that satisfy a weakness indicator while their equivalent www-domains do not, is higher than the number of www-domains that satisfy a weakness indicator while their equivalent plain-domains do not (see \autoref{tab:detailed_diffs_scan2}, under the \quotes{Type/Condition} column, compare the results of row \quotes{$\var{weak}_{\var{plain}} \wedge\lnot \var{weak}_{\var{www}}$} to those of \quotes{$\var{weak}_{\var{www}} \wedge\lnot \var{weak}_{\var{plain}}$} in each dataset). For example, as depicted in \autoref{tab:detailed_diffs_scan2}, in the \texttt{top-domains} dataset, of the 3.62\% plain-domains that provide expired certificates, there are 10.24\% that provide non-expired certificates by their equivalent www-domains. However, in the same dataset (\texttt{top-domains}), of the 3.42\% www-domains that provide expired certificates, there are only 5.18\% that provide non-expired certificates by their equivalent plain-domains. The same trend appears in all of the weakness indicators we study (see \autoref{tab:detailed_diffs_scan2}), which suggests that www-domains tend to have better TLS security configurations and certificates than their equivalent plain-domains. 

\subsection{Relationship to HTTPS Redirection}
To better understand the reasons behind the observed phenomenon, we analyse the HTTPS redirection behaviour of those domains. We input \num{678214} top domains and \num{612839} random domains that have provided responses for both plain-domains and their equivalent www-domains in the TLS~1.3 latest scan experiment. As depicted in \autoref{tab:redirection_scan}, from the \texttt{top-domains} dataset, \num{97.55\%} of plain-domains and \num{97.68\%} of www-domains responded to our redirection scan. From the \texttt{random-domains} dataset, \num{95.92\%} of plain-domains, and \num{96.01\%} of www-domains have responded to our redirection scan. Of the responding domains to our redirection scan, in the \texttt{top-domains} dataset, \num{34.14\%} of HTTPS plain-domains redirected to (i.e. land on) their equivalent HTTPS www-domains, while \num{23.54\%} of HTTPS www-domains redirected to their equivalent HTTPS plain-domains. In the \texttt{random-domains} dataset, \num{13.35\%} of HTTPS plain-domains redirected to their equivalent HTTPS www-domains, while 10.13\% of HTTPS www-domains redirected to their equivalent HTTPS plain-domains. From these results, we conclude that plain-domains tend to redirect to their equivalent www-domains more than www-domains that redirect to their equivalent plain-domains, and HTTPS redirection from plain-domains to their equivalent www-domains is more utilised in \texttt{top-domains} than in \texttt{random-domains}. 

\begin{table}[!tp]
	\centering
	\caption{Summary of the HTTPS redirection scan results.}
	\label{tab:redirection_scan} 
	\begin{adjustbox}{max width=\columnwidth}
		\begin{tabular}{lllr@{\hspace{3pt}}rr@{\hspace{3pt}}r}
			\toprule
			\thead{Dataset} & \thead{Size} &\thead{Type} & \multicolumn{2}{c}{\thead{Responses}} & \multicolumn{2}{c}{\thead{Redirection}}\\
			\midrule
			\multirow{2}{*}{\texttt{top-domains}} & \multirow{2}{*}{\num{678214}} 
			&plain & \num{661596}&(97.55\%) &\num{225839}&(34.14\%) \\
			\cline{3-7}
			& & www  &  \num{662474}&(97.68\%) & \num{155921}&(23.54\%) \\ 
			\midrule
			
			\multirow{2}{*}{\texttt{random-domains}} &   \multirow{2}{*}{\num{612839}} 	   
			&plain &  \num{587850}&(95.92\%) &  \num{78449}&(13.35\%)  \\
			\cline{3-7}
			& & www  & \num{588380}&(96.01\%) & \num{59611}&(10.13\%)   \\
			\bottomrule
		\end{tabular}
	\end{adjustbox}
\end{table}
We conduct further analysis on the set of domains that showed at least one weakness indicator in plain-domains, but not in their equivalent www-domains in the TLS scan (see \autoref{tab:detailed_diffs_scan2}, domains in row \quotes{$\var{weak}_{\var{plain}} \wedge\lnot \var{weak}_{\var{www}}$} in both datasets). In the \texttt{top-domains} dataset, out of \num{11893} top domains that fall into this set and responded to our plain-domains redirection scan, \num{6345} (53.35\%) HTTPS plain-domains redirected to (land on) their equivalent HTTPS www-domains. Of those, \num{1568} (24.71\%) have one or more HTTP (plain-text, unencrypted and unauthenticated) intermediate URLs in the redirection chain. In the \texttt{random-domains} dataset, out of the \num{3369} random domains that fall into that set and responded to our redirection scan, \num{664} (19.71\%) HTTPS plain-domains redirected to their equivalent HTTPS www-domains. Of those, there are \num{252} (37.95\%) HTTP intermediate URLs. The redirection from HTTPS plain-domains to HTTPS www-domains in these domains that show one or more weakness indicators in plain-domains, but not in www-domains, is higher than the overall redirection rate that is presented in \autoref{tab:redirection_scan} earlier. Apparently, in this set of domains, domain administrators pay more attention to www-domains security as HTTPS plain-domains are redirected to their equivalent HTTPS www-domains. However, secure redirection should maintain secure TLS configurations and certificates at every point in the redirection chain.

\section{Related Work}
\label{sec:related}
Chang et al.~\cite{chang17} conducted a study that assessed the HTTPS redirection in top domains. They found that the majority (83.3\%) of HTTPS redirections are not secure. They examine application level properties such as the \texttt{\justify Strict-Transport-Security} (HSTS) policy adoption. Our work looks at the problem from a different layer. That is, we consider transport layer security configurations and certificates, which have not been considered previously. W{\"a}hlisch et al.~\cite{wahlisch15} noticed differences in the IP address prefixes between plain-domains and www-domains as part of their empirical analysis of the Resource Public Key Infrastructure (RPKI) deployment on the \texttt{Alexa} list of the top one million domains. They pointed out that the differences in the top 100K domains are higher than those in the rest of the domains. Our work examines two different datasets, \texttt{top-domains}, and \texttt{random-domains}. Additionally, it compares the TLS configurations and certificates, which have not been explored in previous work. Finally, there are several studies that evaluate the cryptographic strengths of TLS servers such as Lee et al.~\cite{lee07}, Holz et al.~\cite{holz11}, Alashwali~\cite{alashwali13}, and more recently, Alashwali et al.~\cite{alashwali19fs}, Calzavara et al.~\cite{calzavara19}, and Kotzias et al.~\cite{kotzias18}. However, none of them have attempted to compare the TLS configurations when connecting to plain-domains and their equivalent www-domains.

\section{Conclusion}
\label{sec:conclusion}
In this paper, we presented the results of an experiment that aims to
explore whether plain-domains and their equivalent www-domains
differ in TLS security configurations and certificates? and if so,
to what extent? Our results inform the Internet measurement-based
research community, domains (servers) administrators, developers,
and users alike. First, we provided evidence that there is a difference between
plain-domains and their equivalent www-domains in terms of
TLS security configurations and certificates in a non-trivial number
of cases. The difference is more notable in top domains than in random
domains. Second, by defining some weakness indicators and examining when plain-domains and their equivalent www-domains satisfy these indicators, we showed that www-domains tend to have better TLS security than their equivalent
plain-domains. Third, we showed that HTTPS redirection from
plain-domains to their equivalent www-domains is widely utilised, especially in the top domains, and more significantly in the set of domains
that show one or more weakness indicator in plain-domains
but not in their equivalent www-domains ($\var{weak}_{plain} \wedge\lnot \var{weak}_{www}$). In the latter case, users see the final www-domains (URL) with strong TLS security configurations and certificates, but in fact, the HTTPS request has actually passed through plain-domains which have less secure TLS configurations and certificates at previous points in the redirection chain. Even worse, many intermediate URLs in these redirection chains are plain-text HTTP. This introduces a weak link in the system, which may be dangerous for users.

\begin{acks}
We thank the CS's IT and OxCERT teams at the University of Oxford, and the \texttt{tls-scan} developer, Binu Ramakrishnan, for technical support. Pawel's work was supported by the SUTD SRG ISTD 2017 128 grant.
\end{acks} 

\Urlmuskip=0mu plus 1mu\relax
\bibliographystyle{ACM-Reference-Format}
\bibliography{ref}

\end{document}